\documentclass[12pt]{article}

\begin{document}
\setcounter{page}{1}
\renewcommand{\thefootnote}{\fnsymbol{footnote}}
\pagestyle{plain} \vspace{1cm}
\begin{center}
\large{\bf Planck-scale nonthermal correlations in a noncommutative
geometry inspired Vaidya black hole }

\small \vspace{1cm} {\bf S. Hamid
Mehdipour\footnote{mehdipour@liau.ac.ir}}
\\
\vspace{0.5cm} {\it Islamic Azad University, Lahijan Branch,
P. O. Box 1616, Lahijan, Iran }\\

\end{center}
\vspace{1.5cm}

\begin{abstract}
Using the noncommutative geometry inspired Vaidya metric obtained in
terms of coordinate coherent states and also utilizing the
generalized uncertainty principle (GUP), we show that the nonthermal
nature of the Hawking spectrum leads to Planck-scale nonthermal
correlations between emitted modes of evaporation. Our analysis thus
exhibits that owing to self-gravitational effects plus
noncommutativity and GUP influences, information can emerge in the
form of Planck-scale correlated emissions from the black hole.
\end{abstract}
\newpage

\section{\label{sec:1}Introduction}
The discovery of black hole evaporation from the Hawking proposal
\cite{haw1} has led to a long-lived discussion dealing with the
so-called {\it black hole information paradox} (for reviews see
\cite{pre}) which implies the purely thermal essence of the
spectrum. It is widely believed that a reply to the information loss
problem could provide a key element in the quest for a yet to be
formulated theory of quantum gravity. In quantum mechanics,
information is preserved by unitarity. While, in black hole physics,
the information conveyed by a physical procedure descends to the
singularity, and therefore has never been retrieved. This would
allow a nonunitary quantum evolution that maps a pure state to a
mixed state \cite{haw2}. The reply to the query of whether or not
this contravention transpires will profoundly describe the process
in which quantum mechanics and general relativity (e.g., black hole
physics) would appear as borders of the quantum gravity hypothesis.

There are many conjectures to preserve the fundamental principles of
quantum mechanics near black holes \cite{pre}. A common conjecture
is that information actually comes out of the black hole as
nonthermal correlations between different modes of radiation during
evaporation. The Hawking effect can be described as the quantum
tunneling of vacuum fluctuations across the horizon. This is
achieved either by the radial null geodesic approach \cite{par1} or
by the Hamilton-Jacobi approach \cite{sha} to compute the tunneling
probability. In these methods, the credibility of ordinary
perspective for the radiation procedure will fail owing to the fact
that, in the final phase of black hole evaporation, the effects of
the gravitational back-reaction should be taken into consideration.
In Parikh-Wilczek tunneling \cite{par1}, one takes into account the
back-reaction consequences in a definite gap between the initial and
final radii as a result of self-gravitation effects of outgoing
shells, which is the classically forbidden path, (i.e., the
barrier). From the other point of view, the tunneling barrier, which
guarantees energy conservation throughout the evaporation is created
by a reduction in the black hole horizon just by the emitted
particle itself which leads to nonthermal corrections to the black
hole radiation spectrum. However, the authors of
Refs.~\cite{par2,arz} (incorrectly) show that the nonthermal
spectrum by itself does not solve the information loss problem and
the form of the corrections, due to the absence of correlations
between the tunneling probabilities of different modes in the black
hole radiation spectrum, is not sufficient by itself to regain the
information. This incorrect claim was corrected by Zhang {\it et
al.} \cite{zhan} (see also \cite{sing}). In 2009, Zhang {\it et
al.}, by using a standard statistical method and based on the
results within a semiclassical treatment for s-wave emissions,
discovered the existence of correlations within Hawking radiation
from a black hole. They demonstrated that black hole radiation as
tunneling is an entropy conservation process, and information leakes
out via radiation, which clearly leads to the conclusion that the
process of Hawking radiation is unitary and therefore no information
loss appears. The mistake in \cite{par2,arz} was a statistical error
driven from statistically independent events. When performing the
method of \cite{par2,arz} in a purely thermal spectrum, correlations
are observable, which is patently untrue. Whereas in the method of
\cite{zhan} no correlations appear for a purely thermal spectrum.

Lately, a new interesting model of noncommutativity in terms of
coherent states is proposed \cite{sma}, which guarantees Lorentz
invariance, unitarity and UV-finiteness of quantum field theory. The
authors in \cite{nic} used this method to establish a physically
inspired type of noncommutativity corrections to black hole
solutions (coordinate coherent state (CCS) approach). In this model,
the point-like structure of mass $M$, in lieu of being completely
localized at a point, is portrayed by a smeared structure throughout
a region of linear size $\sqrt{\theta}$ (see also \cite{riz}). Using
the CCS approach, it has been exhibited that the modified metric
does not allow the black hole to decay below the Planckian relic.
The evaporation process terminates when the size of the black hole
reaches a Planck size remnant, interpreting a black hole released of
curvature singularity in the origin. Because spacetime
noncommutativity can cure some kinds of divergences that appear in
general relativity, we hope to make some improvements in evaporation
process computations and generalize the tunneling picture using the
CCS method. In 2011, Zhang {\it et al.} accomplished some work in
this direction \cite{zhan2}. They discovered correlations that can
carry information about noncommutativity in Hawking radiation from
noncommutative black holes.

Besides, it is by now widely accepted that measurements in quantum
gravity should be determined by the generalized uncertainty
principle (GUP) \cite{gup1}. In other words, the so-called
Heisenberg uncertainty principle (HUP), should be re-formulated
owing to the noncommutative nature of spacetime at the Planck scale.
As a result, it has been pointed out that in quantum gravity there
exists a minimal observable distance of the order of the Planck
length, which is an immediate consequence of the GUP. Because
quantum gravity proposals prevalently anticipate the existence of a
minimal observable length of the order of the Planck length
\cite{gup1}, the application of the GUP to black hole thermodynamics
has attracted considerable attention and leads to significant
modifications to the emission process, particularly at the final
stages of evaporation (many authors considered various problems in
this framework, e.g. see \cite{gup2}). Recently, we have modified
the Parikh-Wilczek tunneling methodology by including quantum
gravity effects that were revealed in the existence of a minimal
observable length \cite{meh1,meh2,faz,niro}. Indeed, the
self-gravitation influences with inclusion of Planck-scale
modification cannot be neglected, particularly once the black hole
mass becomes comparable to the Planck mass.

In the study of black hole evaporation, there has been a significant
point raised concerning how black hole mass reduces as a
back-reaction of the Hawking radiation. Because the dynamics for the
mass of an evaporating black hole is a persistent problem, we apply
the noncommutative geometry inspired Vaidya metric derived in
Refs.~\cite{mehdi1,mehdi2} to find the Planck-scale nonthermal
correlations within the Hawking radiation. We investigate the
tunneling methodology by the radial null geodesic approach in the
background of CCS noncommutativity including Planck-scale
corrections from the GUP origin. When the effects of gravitational
back-reaction including CCS noncommutativity are incorporated with
the Planck-scale corrections via the GUP, one would recognize the
occurrence of Planck-scale correlations between the tunneling
probability of different modes in the black hole radiation spectrum.
The appearance of these correlations can shed more light on the
information loss problem.

The paper is organized as follows. In Sec.~\ref{sec:2}, using the
influences of noncommutativity in the context of CCS for the Vaidya
metric, we take into account the GUP effects, an achievable role of
quantum gravity, in the Parikh-Wilczek tunneling method. The
tunneling amplitude at which massless particles tunnel across the
event horizon is calculated and the results, a nonthermal spectrum
for the escape of information via the Hawking radiation, is
exhibited, namely the appearance of Planck-scale nonthermal
correlations. Finally, a summary is presented in Sec.~\ref{sec:3}.

\section{\label{sec:2}Noncommutativity and GUP influences on a Vaidya black hole}
According to \cite{nic,riz}, the simple idea of point-like particles
turns into a physically irrelevant concept and should be replaced by
a gaussian mass or energy distribution with a minimal width that
corresponds to the principles of quantum mechanics. The procedure we
use here is to seek for a nonstatic, spherically symmetric,
asymptotically flat structure with a minimal width and gaussian
distribution of mass or energy, whose noncommutative size is
characterized by the parameter $\sqrt{\theta}$. To this purpose, the
mass or energy distribution can be written as
\begin{equation}
\label{mat:1}\rho_{\theta}={M\over
{(4\pi\theta)^{\frac{3}{2}}}}e^{-\frac{r^2}{4\theta}},
\end{equation}
where, $\rho_{\theta}$ and $M$ are functions of both $t$ and $r$.
Because a nonstatic and spherically symmetric spacetime is
contingent upon an arbitrary dynamical mass function, it may be
suitably exhibited by the Vaidya solution \cite{vai,lind}. This kind
of black hole is considered the illustration of a more practical one
due to its time-dependent decreasing mass on account of the
evaporation procedure. In this paper, we use the diagonal form of
the Vaidya metric with respect to $\{x^\mu\}=\{t, r, \vartheta,
\phi\}$ coordinates, ($\mu=0,1,2,3$), as given by Farley and D'Eath
\cite{far}{\footnote{See also \cite{sia} for a more detailed study
of the semiclassical methods which lead to the Hawking temperature
in the Vaidya black hole.}}
\begin{equation}
\label{mat:1.1}ds^2=-e^{b(t,r)}dt^2+e^{a(t,r)}dr^2+r^2d\Omega^2,
\end{equation}
where $d\Omega^2 = d\vartheta^2 + sin^2\vartheta \,d\phi^2$. The
Einstein field equations $G_{\mu\nu}=8\pi T_{\mu\nu}$, for a
spherically symmetric geometry of (\ref{mat:1.1}) lead to the
following relations\footnote{We use the natural units, i.e. $ \hbar=
c = G = 1$.}:
\begin{equation}
\label{mat:1.2}a'-\frac{1-e^a}{r}=8\pi rT_{rr},
\end{equation}
\begin{equation}
\label{mat:1.3}b'+\frac{1-e^a}{r}=8\pi re^{a-b}T_{tt},
\end{equation}
\begin{equation}
\label{mat:1.4}T_{rr}=e^{a-b}T_{tt},
\end{equation}
\begin{equation}
\label{mat:1.4.1}\dot{a}=8\pi rT_{tr},
\end{equation}
where the prime abbreviates $\partial/\partial r$, and the overdot
abbreviates $\partial/\partial t$. Using these equations, one can
find
\begin{equation}
\label{mat:1.5}\frac{a'-b'}{2}=\frac{1-e^a}{r}.
\end{equation}
Now, corresponding to a spherically symmetric null-fluid source, we
can conclude the following expressions for $e^{-a}$ and $e^{b}$:
\begin{equation}
\label{mat:1.6}e^{-a(t,r)}=1-\frac{2M(t,r)}{r},
\end{equation}
\begin{equation}
\label{mat:1.7}
e^{b(t,r)}=\left(\frac{\dot{M}}{\chi(M)}\right)^2e^{-a},
\end{equation}
where $\chi(M)=M'\left(1-2M/r\right)$ is the arbitrary positive
function of $t$ and $r$. We now relate the diagonal form of the
Vaidya metric, as given in Eq.~(\ref{mat:1.1}), to the null form of
the metric. The Vaidya null dust collapse model is described by a
spherically symmetric, nonstatic spacetime with a metric in terms of
null coordinates $(u, r, \vartheta, \phi)$. The Vaidya null dust
collapse is extensively studied in the literature \cite{josh}. In
2001, Claudel, Virbhadra, and Ellis \cite{clau} have done some very
important work in this area. They presented a comprehensive paper
about the geometry of photon surfaces and proved some important
theorems on photon sphere. They have shown that the naked central
singularity for a Vaidya null dust collapse is enclosed within the
photon surface in the sense that any partial Cauchy surface
extending to spatial infinity must intersect the photon surface in a
two-sphere (to perceive implications of photon spheres for
astrophysics, see \cite{virb1,virb2,virb3}).

In the $(u, r, \vartheta, \phi)$ coordinate system, the Vaidya null
dust collapse model has the following form
\begin{equation}
\label{mat:1.8}ds^2=-\left(1-\frac{2M(u)}{r}\right)du^2+2dudr+r^2d\Omega^2.
\end{equation}
It is clear that (\ref{mat:1.8}) is of the Eddington-Finkelstein
type \cite{misn}. The radially outgoing null geodesics are exactly
paths of constant $u$. The function $M$ is now independent of $r$
and constant along outgoing null rays. In the general situation that
$dM/du$ is not known, it has demonstrated unfeasible to diagonalise
the Vaidya metric and to determine $u$ as an explicit function of
$t$ and $r$. In view of the fact that $\dot{M}<0$ and $M'>0$, one
obtains that, along lines: ${u = constant}$, $r$ increases with
increasing $t$. An alteration of variables: $(u, r)\rightarrow(t,
r)$, which agrees asymptotically with the requirement $u = (t - r)$,
can be found by the coordinate transformation \cite{far}
\begin{equation}
\label{mat:1.9}du=-\left(\frac{\dot{M}}{\chi(M)}\right)dt-\left(\frac{M'}{\chi(M)}\right)dr.
\end{equation}
To have an exact $(t-r)$ dependent case of the metric, we present a
Schwarzschild-like metric for the Vaidya solution instead of a
standard Eddington-Finkelstein metric. In the following, to make the
problem well-behaved, we choose $\chi(M) = -\dot{M}$. This metric
resembles the Schwarzschild spacetime, excluding that the role of
the Schwarzschild mass is performed by a mass function $M(t, r)$,
which changes extremely gradually with respect to both $t$ and $r$
in the spacetime region including the outgoing radiation \cite{ste}.
Hence, the corresponding geometry in this area containing the
radially outgoing radiation is of a slowly varying Vaidya type, that
is, $\dot{M}\ll 1$ and $M'\ll 1$.

According to \cite{mehdi1,mehdi2}, the noncommutative geometry
inspired Vaidya metric in the presence of a smeared mass or energy
source, by solving Einstein equations with (\ref{mat:1}) as a matter
source, can be found as
\begin{equation}
\label{mat:2}ds^2=-F(t,r)dt^2+F^{-1}(t,r)dr^2+r^2d\Omega^2,
\end{equation}
with
\begin{equation}
\label{mat:3}F(t,r)=1-\frac{2M_\theta(t,r)}{r},
\end{equation}
where the gaussian-smeared mass distribution is
\begin{equation}
\label{mat:4}M_\theta(t,r)=M_I\left[
{\cal{E}}\left(\frac{r-t}{2\sqrt{\theta}}\right)\left(1+\frac{t^2}{2\theta}\right)
-\frac{r}{\sqrt{\pi\theta}}e^{-\frac{(r-t)^2}{4\theta}}\left(1+\frac{t}{r}\right)\right],
\end{equation}
where $M_I$ is the initial black hole mass and ${\cal{E}}(x)$
displays the {\it Gauss error function} specified as
${\cal{E}}(x)\equiv 2/\sqrt{\pi}\int_{0}^{x}e^{-p^2}dp$. Line
element (\ref{mat:2}) portrays the geometry of a noncommutative
inspired Vaidya black hole. It is obvious that metric (\ref{mat:2})
has a coordinate singularity at the event horizon as
\begin{equation}
\label{mat:7}r_H = 2M_\theta(t,r_H).
\end{equation}
Note that because there is no analytical solution for $r_H$ versus
$M_I$, one can approximately compute the noncommutative horizon
radius versus the initial mass by setting $r_H=2M_I$ into the
function of gaussian-smeared mass distribution $M_\theta(t,r_H)$,
namely
\begin{equation}
\label{mat:7.1}r_H=2M_\theta(t,M_I)=2M_I\left[
{\cal{E}}\left(\frac{2M_I-t}{2\sqrt{\theta}}\right)\left(1+\frac{t^2}{2\theta}\right)
-\frac{2M_I}{\sqrt{\pi\theta}}e^{-\frac{(2M_I-t)^2}{4\theta}}\left(1+\frac{t}{2M_I}\right)\right].
\end{equation}

The radiating property of such a modified vaidya black hole can now
be inspected by the quantum tunneling procedure proposed in
Ref.~\cite{par1}. To describe the quantum tunneling approach wherein
a particle travels in a dynamic geometry and crosses the horizon
without singularity on the path, we should use a coordinate system
that is not singular at the horizon. Painlev\'{e} coordinates
\cite{pai} which are utilized to remove coordinate singularity are
specifically appropriate choices in this method. Under the
Painlev\'{e} time coordinate transformation, we have
\begin{equation}
\label{mat:9} dt\rightarrow dt-\frac{\sqrt{1-F(t,r)}}{F(t,r)}dr,
\end{equation}
the noncommutative Painlev\'{e} metric now immediately reads
$$ds^2=-F(t,r)dt^2+2\sqrt{1-F(t,r)}dtdr+dr^2+r^2d\Omega^2$$
\begin{equation}
\label{mat:10}=-\left(1-\frac{2M_\theta(t,r)}{r}\right)dt^2+2\sqrt{\frac{2M_\theta(t,r)}{r}}dtdr+dr^2+r^2d\Omega^2.
\end{equation}
This metric is stationary, and there exists no coordinate
singularity at the horizon. The outgoing radial null geodesics are
given by
\begin{equation} \label{mat:11}
\dot{r}=1-\sqrt{1-F(t,r)}=1-\sqrt{\frac{2M_\theta\left(t,r\right)}{r}},
\end{equation}
where $\dot{r}\equiv dr/dt$. In accordance with the original work by
Parikh and Wilczek \cite{par1}, the WKB approximation is valid at
the neighborhood of the horizon. Therefore, the tunneling
probability for the classically prohibited area as a function of the
imaginary part of the action for a particle in a tunneling procedure
takes the form {\footnote {Note that there is another standpoint on
using Eq.~(\ref{mat:12}); there exists a problem here recognized as
{\it the factor $2$ problem} \cite{akh,cho,pil}. In
Ref.~\cite{akh2}, a solution to this problem was prepared concerning
the overlooked temporal contribution to the tunneling amplitude.}}
\begin{equation}
\label{mat:12}\Gamma\sim e^{-2\textmd{Im}\,\textit{I}}.
\end{equation}
Here, we consider a spherical positive energy shell including the
ingredients of massless particles each of which moves on a radial
null geodesic like an $s$-wave outgoing particle that passes through
the horizon in the outward direction from $r_{in}$ to $r_{out}$. So,
the imaginary part of the action is given by
\begin{equation}
\label{mat:13}\textmd{Im}\,
I=\textmd{Im}\int_{r_{in}}^{r_{out}}p_rdr=\textmd{Im}\int_{r_{in}}^{r_{out}}\int_0^{p_r}dp'_rdr.
\end{equation}
If we consider the particle's self-gravitation effect, in conformity
with the original work suggested by Kraus and Wilczek \cite{kra},
then both the noncommutative Painlev\'{e} metric and the geodesic
equation should be modified by the response of the background
geometry.

As indicated briefly in the introduction, a consequential
anticipation of various scenarios of quantum gravity is the
existence of a minimal observable distance on the order of the
Planck length that cannot be probed \cite{gup1}, for example, in
string theory there exists a constraint on probing distances smaller
than the string length. Therefore the HUP is modified to incorporate
this constrained resolution of spacetime points. The consequence of
this modification is the so-called GUP, which in fact has its origin
in the quantum fluctuations of the spacetime at the Planck scale.
The form of the GUP in terms of the Planck length, $L_{Pl}$, can be
represented as follows:
\begin{equation}
\label{mat:14}\Delta x\geq\frac{1}{\Delta p}+\alpha L_{Pl}^{2}\Delta
p,
\end{equation}
where $\alpha$ is a dimensionless constant on the order of one that
depends on the details of quantum gravity theory. In the limit
$\Delta x\gg L_{Pl}$, the HUP is recovered (i.e., $\Delta x\Delta
p\geq1$). The second term on the right-hand side of the GUP relation
plays an essential role when the momentum and distance scales are in
the vicinity of the Planck scale. In an innovative method, by
applying the HUP, the thermodynamical quantities for a spherical
black hole can be achieved \cite{hup}. Also, the application of the
GUP to black hole thermodynamics in the same method modifies the
results by inclusion of quantum gravity influences on the ultimate
phases of the evaporation process with an abundant phenomenology
\cite{gup2}.

Here, we apply the GUP to find the Planck-scale information in the
black hole evaporation procedure. In this setup, we use the method
appearing in Ref.~\cite{zhan} to recover information from the
Hawking radiation. We are going to investigate the modifications of
the Hawking radiation via the tunneling process by using the
GUP-corrected de Broglie wavelength, the squeezing of the
fundamental momentum-space cell (see for instance \cite{ame} and
references therein), and then a GUP-corrected energy
\begin{equation}
\label{mat:15}\lambda\simeq\frac{1}{p}\left(1+\alpha
L_{Pl}^2p^2\right),
\end{equation}
\begin{equation}
\label{mat:16}\texttt{E}\simeq E(1+\alpha L_{Pl}^2E^2).
\end{equation}
Now, in the tunneling process, it is necessary to take into account
the reaction of the background geometry with an emitted
GUP-corrected energy $\texttt{E}$. We hold the total ADM mass
($M_I$) of the spacetime fixed, and allow the hole mass to
fluctuate. In other words, a massless particle as a shell travels on
the geodesics of a spacetime with $M_I$ replaced by
$M_I-\texttt{E}$. Next, we should first substitute $M_I -
\texttt{E}$ for $M_I$ in Eq.~(\ref{mat:11}) and then apply the
deformed Hamilton's equation of motion \cite{meh1,meh2,faz},
\begin{equation}
\label{mat:17}\dot{r}\simeq\left(1+\alpha L_{Pl}^2
\texttt{E}^2\right)\frac{dH}{dp_r}\Big|_r,
\end{equation}
to alter the integral variable of the imaginary action
(\ref{mat:13}) from momentum to energy. So, we have
\begin{equation}
\label{mat:18}\textmd{Im}\,
I=\textmd{Im}\int_{r_{in}}^{r_{out}}\int_{M_I}^{M_I-{\texttt{E}}}\frac{1+\alpha
L_{Pl}^2\texttt{E}'^2}{\dot{r}}dHdr,
\end{equation}
where the hamiltonian is $H = M_I - \texttt{E}'$. We evaluate
integral (\ref{mat:18}) by writing the explicit form for the radial
null geodesic, which includes back-reaction effects, namely
\begin{equation}
\label{mat:19}
\dot{r}=1-\sqrt{\frac{2M_\theta\left(t,M_I-\texttt{E}\right)}{r}},
\end{equation}
where
$$M_\theta\left(t,M_I-\texttt{E}\right)=(M_I-\texttt{E})\Bigg[
{\cal{E}}\left(\frac{2(M_I-\texttt{E})-t}{2\sqrt{\theta}}\right)\left(1+\frac{t^2}{2\theta}\right)$$
\begin{equation}
\label{mat:20}-\frac{2(M_I-\texttt{E})}{\sqrt{\pi\theta}}e^{-\frac{(2(M_I-\texttt{{\tiny{E}}})-t)^2}{4\theta}}\left(1+\frac{t}{2(M_I-\texttt{E})}\right)\Bigg].
\end{equation}
Thus, we find
\begin{equation}
\label{mat:21}\textmd{Im}\,
I=\textmd{Im}\int_{r_{in}}^{r_{out}}\int_{0}^{{\texttt{E}}}\frac{1+\alpha
L_{Pl}^2\texttt{E}'^2}
{1-\sqrt{\frac{2M_\theta\left(t,M_I-\texttt{E}'\right)}{r}}}(-d{\texttt{E}'})dr.
\end{equation}
The $r$ integral of Eq.~(\ref{mat:21}) can be done first by
deforming the contour for the lower half $\texttt{E}'$ plane.
Finally, the imaginary part of the action yields the following form:
\begin{equation}
\label{mat:22}\textmd{Im}\,I=\textmd{Im}\int_{0}^\texttt{E}4\pi
iM_\theta(t,M_I-\texttt{E}')\left(1+\alpha L_{Pl}^2
\texttt{E}'^2\right)d\texttt{E}',
\end{equation}
which can be achieved as
$$\textmd{Im}\,I=3\pi\theta({\cal{E}}_2-{\cal{E}}_1)+\sqrt{\pi\theta}e^{-u}\big[(6M_I+5t)(e^v-e^w)+6Ee^w\big]+\pi\alpha L_{Pl}^2\Bigg[
({\cal{E}}_2-{\cal{E}}_1)$$$$\times\bigg(\frac{15}{16}t^4-\frac{10}{3}M_It^3+3M_I^2t^2-\frac{M_I^4}{3}\bigg)+{\cal{E}}_2E^3\left(\frac{16}{3}M_I-5E\right)
\Bigg]+\pi\Big[({\cal{E}}_2-{\cal{E}}_1)$$$$\times\left(3t^2-2M_I^2\right)+2{\cal{E}}_2E(2M_I-E)\Big]+\sqrt{\frac{\pi}{\theta}}t^2e^{-u}\Bigg[\alpha
L_{Pl}^2\Bigg[e^v\Bigg(\frac{t^3}{16}+\frac{M_It^2}{24}-\frac{M_It}{8}$$$$-\frac{M_I^2}{12}\Bigg)-e^w\Bigg[\frac{t^3}{16}-t^2\left(\frac{5M_I}{24}
+\frac{E}{8}\right)+t\left(\frac{M_IE}{6}+\frac{M_I^2}{12}+\frac{E^2}{4}\right)+\frac{EM_I^2}{6}-\frac{3E^3}{2}$$$$+\frac{M_I^3}{6}+\frac{M_I
E^2}{6}\Bigg]\Bigg]+\left(M_I+\frac{t}{2}\right)(e^v-e^w)+Ee^w\Bigg]+\frac{\pi}{\theta}\Bigg[\alpha
L_{Pl}^2\Bigg[({\cal{E}}_2-{\cal{E}}_1)\bigg(\frac{t^6}{32}-\frac{M_It^5}{6}$$
$$+\frac{M_I^2t^4}{4}-\frac{M_I^4t^2}{6}\bigg)+t^2E^3{\cal{E}}_2\left(
\frac{8M_I}{3}-\frac{5E}{2}\right)\Bigg]+({\cal{E}}_2-{\cal{E}}_1)\left(\frac{t^4}{4}-M_I^2t^2\right)$$\begin{equation}
\label{mat:23}+{\cal{E}}_2t^2E(2M_I-E)\Bigg]+O(\alpha^2L_{Pl}^4),
\end{equation}
where
\begin{displaymath}
\left\{ \begin{array}{ll}
{\cal{E}}_1\equiv{\cal{E}}\left(\frac{2M_I-t}{2\sqrt{\theta}}\right)\\
{\cal{E}}_2\equiv{\cal{E}}\left(\frac{2M_I-2E-t-2\alpha L_{Pl}^2 E^3}{2\sqrt{\theta}}\right)\\
u\equiv \frac{M_I^2+E^2+Et+\alpha L_{Pl}^2\left(2E^4+E^3t\right)}{\theta}\\
v\equiv u-\frac{t^2+4M_I(M_I-t)}{4\theta}\\
w\equiv \frac{4M_I\left(2E+t+2\alpha
L_{Pl}^2E^3\right)-t^2}{4\theta}.
\end{array} \right.
\end{displaymath}
We consider the leading-order correction to be just proportional to
$(\alpha L_{Pl}^2)$. These new corrections cannot be ignored when
the black hole mass is close to the Planck mass. However, the
corrections are substantially trivial, one could observe this as a
consequence of quantum inspection at the level of semi-classical
quantum gravity. Note that we have eliminated the terms proportional
to $(\alpha L_{Pl}^2\sqrt{\theta})$ and also $(\alpha
L_{Pl}^2\theta)$ owing to their smallness in nature, to preserve the
integrity of Eq.~(\ref{mat:23}), and for one's convenience.

The imaginary part of the action, in high energies, can be written
as \cite{kes,ban,mas}
\begin{equation}
\label{mat:24}\textmd{Im}\,I=-\frac{1}{2}\Delta
S_{NC}=-\frac{1}{2}\left[S_{NC}(M_I-E)-S_{NC}(M_I)\right],
\end{equation}
where $S_{NC}$ is the noncommutative black hole entropy. From this
viewpoint the emission rate is proportional to the difference in
black hole entropies before and after emission which means that the
emission spectrum cannot be accurately thermal at higher energies.
From Eq.~(\ref{mat:23}), it is clearly observed that the
corresponding tunneling amplitude disagrees with the purely thermal
spirit of the spectrum. It can be simply confirmed that the energy
conservation or self-gravitational effect plus the additional or
combined terms depending on the parameters GUP and noncommutativity
(i.e., $\alpha$ and $\theta$, respectively) lead to a Planck-scale
statistical correlation function between probabilities of tunneling
of two particles with different energies, that is,
\begin{equation}
\label{mat:26}C(E_1+E_2;E_1,E_2)=\ln\Gamma(E_1+E_2)-\ln\left[\Gamma(E_1)\Gamma(E_2)\right]
\neq0,
\end{equation}
where
\begin{displaymath}
\left\{ \begin{array}{ll} \Gamma(E_1)=\Lambda\int_0^{M_I-E_1}
\Gamma(E_1,E_2)dE_2\\
\Gamma(E_2)=\Lambda\int_0^{M_I-E_2}
\Gamma(E_1,E_2)dE_1\\
\Gamma(E_1,E_2)= \Gamma(E_1+E_2)\\
\Lambda=\left[ \int_0^{M_I} e^{\Delta S_{NC}}dE\right]^{-1}.
\end{array} \right.
\end{displaymath}
This means that the probability of tunneling of two particles with
energies $E_1$ and $E_2$ is not equal to the probability of
tunneling of one particle with their compound energies $E=E_1+E_2$,
as expected from a nonthermal spectrum \cite{zhan}. Hence subsequent
Hawking radiation emissions must be correlated. It is not necessary
to write the expression $C$ in terms of energies because it is too
long even after simplifying. We have checked our result in some
limits. In the limit $\theta\rightarrow 0$, $t=0$, and
$\alpha\neq0$, one finds
$$C(E_1+E_2;E_1,E_2)=8\pi E_1E_2-2\pi\alpha
L_{Pl}^2\bigg[(16M_I-15(E_1+E_2))(E_1^2E_2+E_1E_2^2)$$\begin{equation}
\label{mat:27}-5(E_1^3E_2+E_1E_2^3)\bigg].
\end{equation}
The existence of an additional term depending on the GUP parameter
on the right-hand side of Eq.~(\ref{mat:27}) is due to nonthermal
GUP correlations. It is evident that in the HUP limit, $\alpha=0$,
with $\theta=0$ and $t=0$, we regain the same result as
Ref.~\cite{zhan}, that is,
\begin{equation}
\label{mat:28}C(E_1+E_2;E_1,E_2)=8\pi E_1E_2.
\end{equation}
In the HUP limit with nonzero $\theta$ and $t=0$, we obtain
$$C(E_1+E_2;E_1,E_2)=4\pi\Bigg(\left[(M_I-E_1-E_2)^2-\frac{3}{2}\theta\right]
{\cal{E}}\left(\frac{M_I-E_1-E_2}{\sqrt{\theta}}\right)$$$$-\left[(M_I-E_1)^2-\frac{3}{2}\theta\right]
{\cal{E}}\left(\frac{M_I-E_1}{\sqrt{\theta}}\right)-\left[(M_I-E_2)^2-\frac{3}{2}\theta\right]
{\cal{E}}\left(\frac{M_I-E_2}{\sqrt{\theta}}\right)$$$$+\left[M_I^2-\frac{3}{2}\theta\right]
{\cal{E}}\left(\frac{M_I}{\sqrt{\theta}}\right)\Bigg)+12\sqrt{\pi\theta}\bigg((M_I-E_1-E_2)e^{-\frac{(M_I-E_1-E_2)^2}{\theta}}-
(M_I-E_1)$$\begin{equation} \label{mat:29}\times
e^{-\frac{(M_I-E_1)^2}{\theta}}-(M_I-E_2)e^{-\frac{(M_I-E_2)^2}{\theta}}+M_Ie^{-\frac{M_I^2}{\theta}}\bigg).
\end{equation}
If one takes the following approximations:
\begin{displaymath}
\left\{ \begin{array}{ll}
{\cal{E}}\left(\frac{2(M_I-E_1-E_2)-t}{2\sqrt{\theta}}\right)\simeq{\cal{E}}\left(\frac{2(M_I-E_1)-t}{2\sqrt{\theta}}\right)\simeq
{\cal{E}}\left(\frac{2(M_I-E_2)-t}{2\sqrt{\theta}}\right)\simeq{\cal{E}}\left(\frac{2M_I-t}{2\sqrt{\theta}}\right)\\
e^{-\frac{(2(M_I-E_1-E_2)-t)^2}{4\theta}}\simeq
e^{-\frac{(2(M_I-E_1)-t)^2}{4\theta}}\simeq
e^{-\frac{(2(M_I-E_2)-t)^2}{4\theta}}\simeq
e^{-\frac{(2M_I-t)^2}{4\theta}},
\end{array} \right.
\end{displaymath}
then for $\alpha=0$, we have
\begin{equation}
\label{mat:30}C(E_1+E_2;E_1,E_2)=8\pi
E_1E_2\bigg(1+\frac{t^2}{2\theta}\bigg){\cal{E}}\left(\frac{2M_I-t}{2\sqrt{\theta}}\right).
\end{equation}
Substituting $t=0$ into Eq.~(\ref{mat:30}), we get
\begin{equation}
\label{mat:31}C(E_1+E_2;E_1,E_2)=8\pi
E_1E_2{\cal{E}}\left(\frac{M_I}{\sqrt{\theta}}\right),
\end{equation}
which is directly obtained from Eq.~(\ref{mat:29}) by applying the
approximations given earlier with $t=0$. For the commutative case,
$\frac{M_I}{\sqrt{\theta}}\rightarrow\infty$, the Gauss error
function in Eq.~(\ref{mat:31}) tends to unity and one recovers a
similar result to Ref.~\cite{zhan}. So, the emission rates for
different modes of radiation during evaporation are mutually related
to one another from a statistical viewpoint. Moreover, the inclusion
of the effects of quantum gravity as a GUP expression plus
noncommutativity influences causes the creation of Planck-scale
correlations between the different modes of radiation.

\section{\label{sec:3}Summary}
In the framework of a noncommutative model of CCS, we have
considered a Schwarzschild-like metric for Vaidya solution instead
of standard Eddington-Finkelstein metric to observe an exact $(t -
r)$ dependent case of the metric. In this situation, we have shown
that incorporation of quantum gravity effects, such as GUP, combined
with the noncommutativity influences leads to Planck-scale
correlations between emitted particles. These features reflect the
fact that the information emanates from the black hole as
Planck-scale nonthermal correlations within the Hawking radiation
and this can shed more light on the information problem in black
hole evaporation.

\end{document}